\documentclass[ reprint, aps, prb, two column, showkeys, groupedaddress]{revtex4-1}
\usepackage{graphicx,epsfig}
\usepackage{amssymb}
\usepackage{amsmath}
\usepackage{bm}
\usepackage{textcomp}
\usepackage{soul,xcolor}

\usepackage{color}

\begin{document}
\setcounter{page}{1}
\setstcolor{red}

\title[]{Heterostructure design to achieve high quality, high density GaAs 2D electron system with $g$-factor tending to zero at large hydrostatic pressures}
\author{Yoon Jang \surname{Chung}}
\email{edwinyc@princeton.edu}
\affiliation{Department of Electrical Engineering, Princeton University, Princeton, NJ 08544, USA  }
\author{S. \surname{Yuan}}
\author{Yang \surname{Liu}}
\affiliation{International Center for Quantum Materials, Peking University, Beijing, 100871, China  }
\author{K. W. \surname{Baldwin}}
\author{K. W. \surname{West}}
\author{M. \surname{Shayegan}}
\author{L. N. \surname{Pfeiffer}}
\affiliation{Department of Electrical Engineering, Princeton University, Princeton, NJ 08544, USA  }
\date{\today}

\begin{abstract}

Hydrostatic pressure is a useful tool that can tune several key parameters in solid state materials. For example, the Land\'e $g$-factor in GaAs two-dimensional electron systems (2DESs) is expected to change from its bulk value  $g\simeq-0.44$ to zero and even to positive values under a sufficiently large hydrostatic pressure. Although this presents an intriguing platform to investigate electron-electron interaction in a system with $g=0$, studies are quite limited because the GaAs 2DES density decreases significantly with increasing hydrostatic pressure. Here we show that a simple model, based on pressure-dependent changes in the conduction band alignment, quantitatively explains this commonly observed trend. Furthermore, we demonstrate that the decrease in the 2DES density can be suppressed by more than a factor of 3 through an innovative heterostructure design.

\end{abstract}
\maketitle

	Clean two-dimensional electron systems (2DESs) have long served as and continue to be the community standard to investigate many-body physics. The main strength of 2DESs is that they can readily be prepared in the laboratory with appropriate parameters so that electron-electron interaction dominates the energy scales in the system. This is especially true when a perpendicular magnetic field is applied to the 2DES because the field quenches the kinetic (Fermi) energy and causes the density of states to collapse into a set of discrete Landau levels. An epitome is the emergence of the fractional quantum Hall effect in the extreme quantum limit, an effect that is completely unexpected in the single particle picture \cite{Tsui,Laughlin,Jain}. The observation of numerous other interaction-driven states such as Wigner solid \cite{Wigner1,Wigner2,Wigner3}, bubble \cite{bubble}, and stripe/nematic \cite{stripe1,stripe2,stripe3} phases are more examples that demonstrate the rich physics embedded in 2DESs under a magnetic field.

	Of course, the characteristics of a many-body phase at a particular magnetic field in a 2DES depend on the energetics of the situation. The typical intricacies involved here are the effective mass ($m^{*}$) and Land\'e $g$-factor ($g$) of the material hosting the 2DES since they are directly tied to the cyclotron and Zeeman energies, respectively. These material parameters derive from the periodic structure of the crystal and the atoms that comprise it, meaning that they are not continuously variable under the usual circumstances. One way to overcome this constraint is to apply hydrostatic pressure to a system as it can gradually modify the lattice constant of the crystal. For example, in GaAs, it has been shown that one can significantly alter $g$ and even tune it to go through zero while leaving $m^{*}$ almost unaffected by applying hydrostatic pressure to the sample \cite{gfactor1,pressure1,pressure2,pressure3,pressure4,pressure5,gfactor2}.
	
	Although this control over tuning the $g$-factor presents exciting opportunities to investigate exotic interaction-driven phases such as skyrmions or quantum Hall ferromagnets \cite{skyrmion,pressure4,pressure5} with very weak or even zero spin splitting in GaAs 2DESs, it suffers from a major drawback: in conventional GaAs 2DESs, it is extremely difficult to sustain the electron density at reasonably high values as hydrostatic pressure is applied \cite{pressure1,pressure2,pressure3,pressure4,pressure5}. In fact, state-of-the-art GaAs 2DESs are expected to be completely depleted at the pressure of $\sim18$ kbar where $g$ would be close to zero for bulk GaAs \cite{gabor}. Here we show that this decrease in 2DES density can be suppressed by more than a factor of 3 through innovation in GaAs 2DES heterostructure design. Our samples display high-quality magnetotransport at large hydrostatic pressures as manifested by the presence of numerous, fractional quantum Hall states, and provide a robust platform to study delicate many-body phenomena under these demanding conditions.
	 
	Before describing our structure, it is important to establish a basic understanding of why the electron density decreases in preexisting GaAs 2DESs under hydrostatic pressure. Figure 1(a) shows a schematic band diagram of a standard modulation-doped structure before the electrons have transferred to the 2DES. Once equilibrated, for a given spacer thickness, the electron density in the 2DES is approximately proportional to the energy difference ($\Delta E_C$) between the ground-state energy of the main GaAs quantum well ($E_0^{main}$) and the donor energy level ($E_D$) \cite{designrules,davies}. This is because after the electrons transfer to the main quantum well, the capacitive energy is the dominant term that comprises $\Delta E_C$ so that $\Delta E_C\simeq E_{cap}=nse^2/\epsilon_b$. Here $n$ is the electron density, $s$ is the spacer thickness, $e$ is the fundamental charge and $\epsilon_b$ is the dielectric constant of the barrier. In this simple structure at atmospheric pressure, the relevant donor level that determines $\Delta E_C$ is tied to the $\Gamma$-band of the AlGaAs barrier, as it is lower in energy compared to the donor level tied to the X-band \cite{footnote1}. However, when hydrostatic pressure is applied to the GaAs/Al$_x$Ga$_{1-x}$As system, the X-band gap decreases while the $\Gamma$-band gap increases, and a $\Gamma$- to X-band crossover occurs for the conduction-band minimum and hence the lowest donor level of the AlGaAs barrier. This is shown quantitatively in Fig. 1(b) for a typical modulation-doped GaAs 2DES with a barrier Al fraction of $x=0.32$, based on parameters available in the literature \cite{Adachi}. The dashed lines in Fig. 1(b) denote the hydrogenic donor levels affiliated with each band (shown by solid lines in Fig. 1(b)); because the effective mass is larger for the X-band compared to the $\Gamma$-band (assuming the density-of-states effective mass, $m_X^*\simeq0.81$ and $m_\Gamma^*\simeq0.093$) \cite{Adachi}, the donor level transition occurs before the actual $\Gamma$- and X-bands cross. The evolution plotted in Fig. 1(b) implies that the GaAs 2DES density using standard modulation-doped structures should show a steady decrease as hydrostatic pressure is applied to the system, which is consistent with previous experimental investigations \cite{pressure1,pressure2,pressure3,pressure4,pressure5,gabor}.
	 
	 \begin{figure} [t]
\centering
    \includegraphics[width=.45\textwidth]{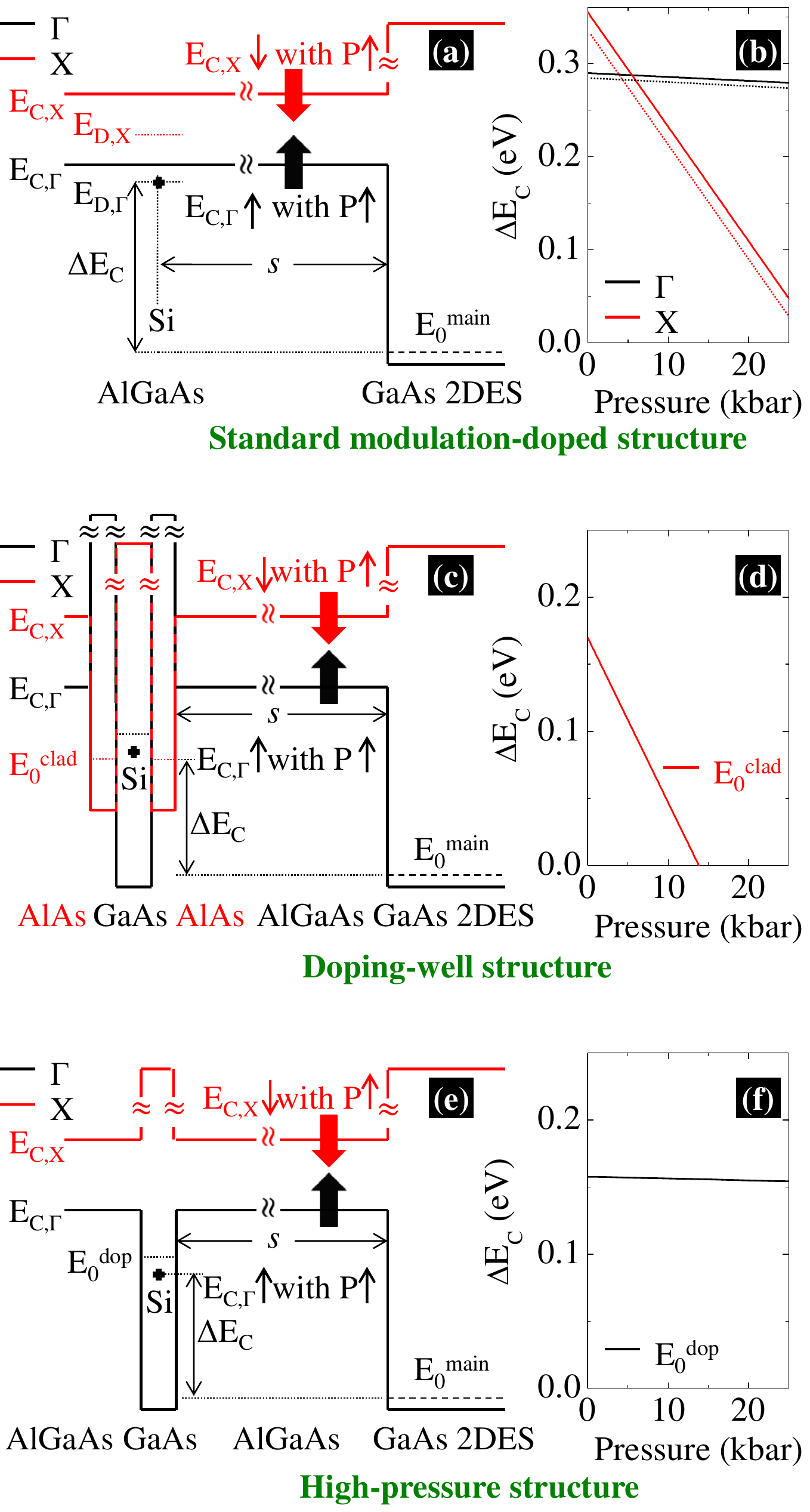}

 \caption{\label{fig1} Schematic band diagram and estimated change in relevant energy levels with respect to $E_0^{main}$ as hydrostatic pressure is applied for: (a), (b) standard modulation doped; (c), (d) doping-well; and (e), (f) our proposed high-pressure structures. For case (c), there is also an AlGaAs layer on the left side of the left AlAs cladding well although it is not specified in the figure due to space limitation. For the sake of simplicity, all bands in (a), (c), and (e), are drawn flat although they should bend in accordance with electric field as charges transfer to the main GaAs 2DES. The values plotted in (b), (d), and (f) were deduced from parameters given in \cite{Adachi}.  }
\end{figure}


	Compared to the standard modulation-doped structure, modern state-of-the-art doping-well structures (DWSs) can host higher quality 2DESs because they provide additional screening from all types of impurities in the sample \cite{Sammon1,Sammon2,dopingwell}. While the DWSs have had a significant impact on common, low-temperature measurements under normal operating conditions, they still suffer from the same issues mentioned above when they are subject to elevated levels of hydrostatic pressure. Shown in Fig. 1(c) is the schematic band diagram of a DWS, where the $\Delta E_C$ that determines the 2DES density is associated with the ground-state energy of an AlAs cladding well ($E_0^{clad}$) rather than a donor level in an AlGaAs barrier \cite{dopingwell}. As delineated earlier, the X-band gap decreases while the $\Gamma$-band gap increases when hydrostatic pressure is applied to the GaAs/AlGaAs system. Similar to the standard modulation-doped case, in the DWS, this causes $\Delta E_C$ and therefore the GaAs 2DES density to decrease as hydrostatic pressure is increased. Figure 1(d) illustrates this in a more elaborate fashion, showing that $\Delta E_C$ decreases linearly with pressure as $E_0^{clad}$ moves with the X-band. Here nearly all the reduction in $\Delta E_C$ comes from the relative change in the band edge of the AlAs cladding well with respect to the main GaAs quantum well because changes in the confinement energy of the cladding well are very small \cite{footnote2}. 
	
	It is evident from the discussion in the previous paragraphs that the depletion of carriers in both standard modulation-doped and DWS GaAs 2DESs under hydrostatic pressure stems from the influence of the X-band on $\Delta E_C$. This implies that, if we can devise a structure in which $\Delta E_C$ is immune to changes in the X-band, we should expect the electron density of the 2DES hosted in it to stay fixed when pressure is applied. Our approach to such a structure is depicted in Fig. 1(e), where $\Delta E_C$ derives from the confinement energy of a narrow GaAs quantum well, instead of an AlAs cladding quantum well as in the case of the DWS. The X-band does not impact $\Delta E_C$ in this design up to very high pressures because the X-band edge is $\simeq0.35$ eV higher than the $\Gamma$-band edge for GaAs. This is better visualized by Fig. 1(f), in which we plot the ground-state energy of the narrow GaAs doping well ($E_0^{dop}$), which determines $\Delta E_C$, as a function of pressure. The slight negative slope in Fig. 1(f) comes from the change in confinement energy of the narrow GaAs quantum well which changes as the barrier height decreases slightly with pressure. Since changes in the confinement energy are more sensitive to the barrier height when the quantum well width is small, this negative slope will become more pronounced the narrower the doping GaAs quantum well is. For the structure used to calculate Fig. 1(f), the doping GaAs quantum well width was 2.83 nm (10 GaAs monolayers), and the rate of change in confinement energy as a function of pressure was $\simeq0.07$ meV/kbar compared to the rate of $\simeq12.8$ meV/kbar for the DWS and $\simeq12.2$ meV/kbar for the standard modulation-doped structure. Even without these numbers, there is a clear distinction in the trend observed in Fig. 1(f) compared to Figs. 1(b) and (d), suggesting that our proposed structure should be able to sustain high 2DES density at large hydrostatic pressures.

\begin{figure}[t]
 
 \centering
    \includegraphics[width=.47\textwidth]{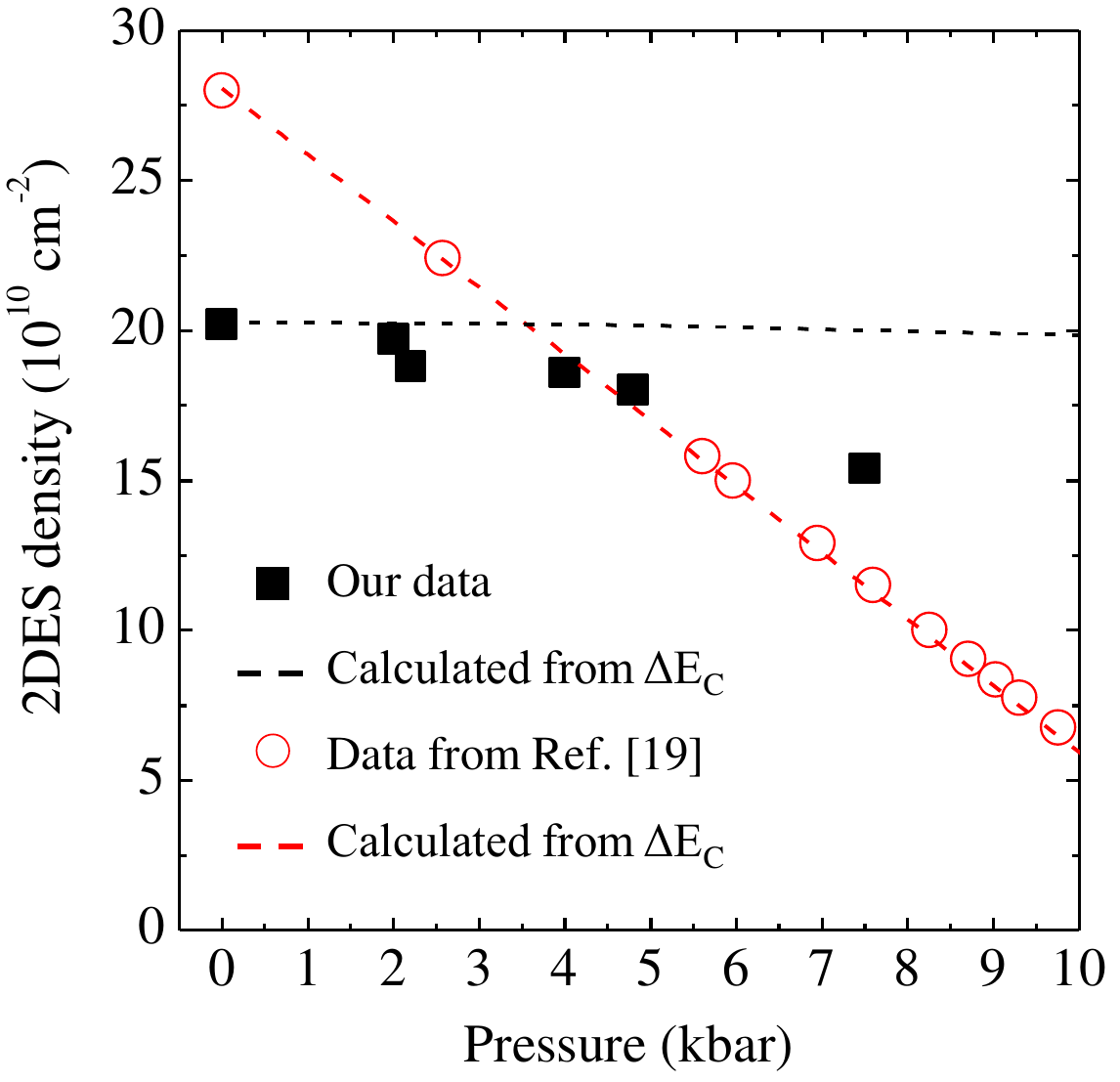} 
  \caption{\label{fig2} Measured 2D electron density vs. hydrostatic pressure for the high-pressure structure presented in this work (solid black squares). Reproduced data from \cite{gabor} for a doping-well structure with similar sample parameters are shown by open red circles for comparison. The black and red dashed lines denote the expected 2D electron density values deduced from the changes in the band energy levels for the high-pressure and doping-well structures, respectively.  }
\end{figure} 

\begin{figure}[t]

\centering
    \includegraphics[width=.47\textwidth]{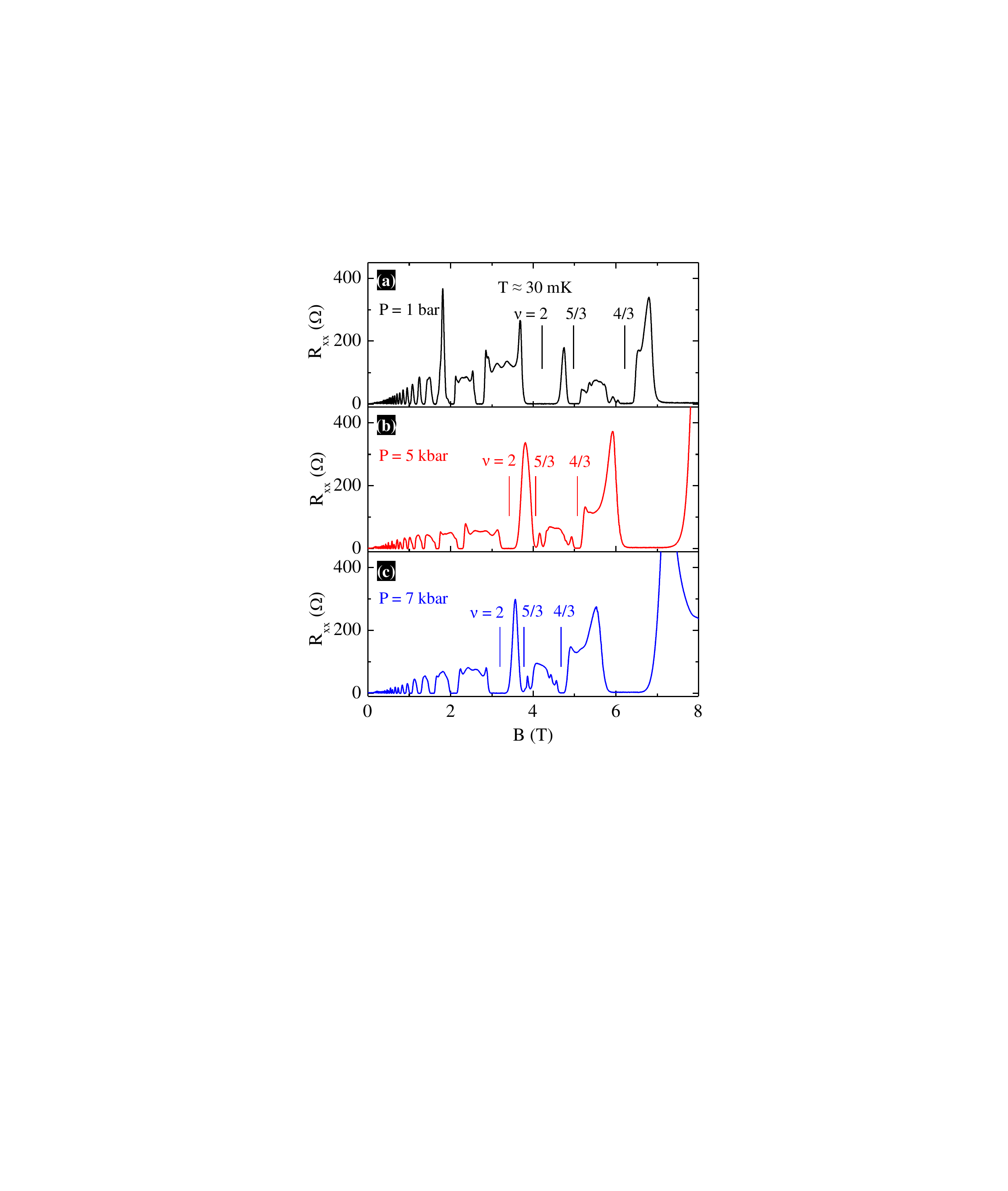} 
  \caption{\label{fig3} Representative magnetotransport traces for our high-pressure sample at pressures of (a) 1 bar, (b) 5 kbar, and (c) 7 kbar; the 2DES densities are 20.1, 16.5, and 15.2$\times10^{10}$ cm$^{-2}$, respectively. All traces were recorded at a temperature of $T\simeq30$ mK. The positions of the Landau level filling factors $\nu$ at which quantum Hall states are observed are marked in each trace.}
\end{figure} 

	We experimentally verified the implication of Fig. 1(f) by growing, via molecular beam epitaxy, the structure of Fig. 1(e), and by measuring the low-temperature ($\sim4$ K) 2DES density at various levels of hydrostatic pressure. Our sample has a spacer thickness ($s$) of 100 nm, main GaAs well width of 30 nm, barrier Al fraction $x=0.38$, and narrow GaAs doping well width of 2.83 nm. The results of our measurements are shown as black squares in Fig. 2. All measurements were performed in the dark. Data points from a previous report using an ultra-high-mobility DWS are shown in red for comparison \cite{gabor}. There is a stark difference in the trends observed for the two data sets, with our proposed structures showing a much slower decrease in density as a function of pressure compared to the DWS. The DWS decreases in density at a rate of $\simeq2.2\times10^{10}$ cm$^{-2}$/kbar while our structures only show a decline of $\simeq6.2\times10^9$ cm$^{-2}$/kbar. This is more than a factor of 3 improvement. It is noteworthy that extrapolating the data out to a pressure of $\sim18$ kbar where bulk GaAs is expected to have $g\simeq0$, our structure would still have an electron density of $\sim1\times10^{11}$ cm$^{-2}$ but the DWS would be fully depleted.

	The dashed lines in Fig. 2 denote the expected electron density deduced from the model discussed in Fig. 1 \cite{footnote3}. As can be seen in Fig. 2, the data for the DWS are in excellent agreement with expected values. For our proposed structure, there appears to be a deviation from our predictions that gets exacerbated at higher pressures. This could be due to factors that were not included in the simple model we present, such as contributions from surface states. However, it is clear from the data that the conduction band alignment issue discussed above is a major contributor to the decrease in 2DES density in previous samples as pressure is applied, and that it can be significantly suppressed by implementing the scheme we propose here. 

	It is also useful to discuss the quality of the magnetotransport data of the proposed sample design. Figures 3(a)-(c) show representative longitudinal resistance ($R_{xx}$) traces of our sample measured at temperature $T\simeq30$ mK at various pressures. The mobility of this sample is $\simeq15\times10^6$ cm$^2$/Vs at $\simeq1$ bar, and the $R_{xx}$ traces exhibit high quality even at high pressures with signatures of numerous fractional quantum Hall states. We would like to note that this is a prototype sample and there is room for further optimization in quality by tailoring the structure for a specific goal.
	
	
	In conclusion, we have designed a GaAs/AlGaAs structure that can sustain a high-quality, high-density 2DES at large hydrostatic pressures. The decrease in the 2DES density as a function of pressure in our sample is suppressed by more than a factor of $3$ compared to state-of-the-art conventional structures. This was achieved by mitigating the conduction band alignment changes that occur between the main quantum well and the doped region as pressure is applied to the system. Our design provides a useful platform to investigate many-body phenomena in 2DESs at extreme conditions.

\begin{acknowledgments}
We acknowledge support through the NSF (Grants DMR 1709076 and ECCS 1906253) for measurements, and the NSF (Grant MRSEC DMR 1420541), Gordon and Betty Moore Foundation (Grant GBMF4420), and the Department of Energy (DOE) Basic Energy Sciences (Grant DE-FG02-00-ER45841) for sample fabrication and characterization.
 \end{acknowledgments}
 
 The data that support the findings of this study are available from the corresponding author upon reasonable request.


\begin{thebibliography}{99}
\bibitem{footnote1} Here we assume hydrogenic donor levels for both the $\Gamma$- and X-bands.
\bibitem{footnote2} The change in the X-band gap of an Al$_x$Ga$_{1-x}$As system when hydrostatic pressure is applied has almost no compositional dependence; see Ref. \cite{Adachi} for details. Since the X-band of the cladding well and its barriers will then move together as pressure is increased, the barrier height will not change for the cladding well and hence we should expect the ground-state energy to show a minimal change.
\bibitem{footnote3} Here $\Delta E_C$ was calculated by performing simple finite barrier calculations for the doping quantum well while considering the conduction band changes as pressure is applied. See Ref. \cite{dopingwell} for more details.
\end{thebibliography}
\end{document}